\documentclass[prd, amsfonts, twocolumn, nofootinbib, showpacs]{revtex4-2}
\usepackage{graphicx, epsfig}
\usepackage{color}
\usepackage{amsmath}
\usepackage{amssymb}
\newcommand{\be}{\begin{equation}}
\newcommand{\ee}{\end{equation}}
\newcommand{\bea}{\begin{eqnarray}}
\newcommand{\eea}{\end{eqnarray}}

\newcommand{\gapp}{\mathrel{\raise.3ex\hbox{$>$}\mkern-14mu
\lower0.6ex\hbox{$\sim$}}}
\newcommand{\lapp}{\mathrel{\raise.3ex\hbox{$<$}\mkern-14mu
\lower0.6ex\hbox{$\sim$}}}
\def\bbox{{\,\lower0.9pt\vbox{\hrule \hbox{\vrule height 0.2 cm
\hskip 0.2 cm \vrule  height 0.2 cm}\hrule}\,}}

\begin{document}
\title{Bulk entropy is crucial to validate the second law of the extended black hole thermodynamics}
\author{De-Chang Dai$^{1,2}$\footnote{communicating author: De-Chang Dai,\\ email: diedachung@gmail.com\label{fnlabel}},
Djordje Minic$^3$ and Dejan Stojkovic$^4$}
\affiliation{$^1$ Center for Gravity and Cosmology, School of Physics Science and Technology, Yangzhou University, 180 Siwangting Road, Yangzhou City, Jiangsu Province, P.R. China 225002 }
\affiliation{ $^2$ CERCA/Department of Physics/ISO, Case Western Reserve University, Cleveland OH 44106-7079}
\affiliation{ $^3$ Department of Physics, Virginia Tech, Blacksburg, VA 24061, U.S.A. }
\affiliation{ $^4$ HEPCOS, Department of Physics, SUNY at Buffalo, Buffalo, NY 14260-1500, U.S.A.}

\begin{abstract}
\widetext

The extended black hole thermodynamics in which the cosmological constant plays the role of pressure significantly enriches the phase structure  of the theory.  
In order to understand the extended black hole thermodynamics more precisely, we let the value of the cosmological constant vary dynamically via tunneling from one vacuum to another 
in a black hole induced vacuum decay.
In this process, entropy of the matter/energy released by a black hole  is crucial  to validate the second law of thermodynamics. In other words, without taking this bulk entropy into account, entropy associated with the black hole and cosmological horizons may not always increase.  Since the bulk entropy is not represented by the black hole and the cosmological horizons, this result calls for a more careful interpretation of the holographic principle in which environmental effects are taken into account. 
\end{abstract}


\maketitle

\section{Introduction and motivation}
Black holes are solutions of Einstein's general relativity that are constantly surprising. While black holes are classically black, in the quantum context they can radiate particles and have a well defined temperature, which allows us to treat them as thermodynamical objects. Bekenstein, Hawking, Bardeen, Carter and others nicely summed up the four laws of black hole thermodynamics  in the 1970s \cite{Bekenstein1972,Bekenstein1973,Hawking1972,Hawking1974,Hawking1975,Bardeen1973}. 

\begin{itemize}

\item The zeroth law: The horizon has a constant surface gravity for a stationary black hole. 

\item The first law: For perturbations of stationary black holes, the change of energy is related to the change of area, angular momentum, and electric charge by
\begin{equation}
dE =\frac{\kappa}{8\pi} dA +\Omega dJ +\Phi dQ
\end{equation}
where $E$ is the internal energy equal to the mass $M$ in the case, $\kappa$ is the surface gravity, $A$ is the horizon area, $\Omega$ is the angular velocity, $J$ is the angular momentum, $\Phi$ is the electrostatic potential and $Q$ is the charge. 

\item The second law: The horizon area is a non-decreasing function of time:
\begin{equation}
\frac{dA}{dt}\geq 0.
\end{equation}

\item The third law: It is impossible to form a black hole with vanishing surface gravity. 

\end{itemize}

These four laws directly correspond to the laws of classical thermodynamics, once the black hole temperature and entropy are assigned as 
\begin{equation}
kT = \frac{\hbar \kappa}{2\pi} \text{, } S=\frac{A}{4\hbar G}
\end{equation} 
where $k$ is the Boltzmann constant. However, these laws were formulated before the discovery of Hawking's radiation. Once the Hawking radiation is included, a black hole can reduce its area and hence reduce its entropy. Then, the second law must be modified to include the entropy of the black hole environment. 
{In principle, the black hole environment depends on a cosmological model, but since we live in a de Sitter (dS) universe it is natural to include a cosmological constant in these laws, even though, in general, the physics of the black hole environment is not only captured by the cosmological constant.} Since there are many models in which the cosmological constant, $\Lambda$, is not a constant but it can change in time (e.g. \cite{Henneaux1984,Caldarelli:1999xj}), we can treat the cosmological constant as a thermodynamical variable. 
A more general black hole first law of thermodynamics with varying cosmological constant was proposed in \cite{Sekiwa:2006qj,Kastor:2009wy} as
\begin{equation}
d M = TdS +\Omega d J +\Phi dQ + V dP .
\end{equation}
Here, $P=\frac{-\Lambda}{8\pi G}$ and $V=\frac{4\pi r_h^3}{3}$, where $r_h$ is the black hole horizon radius. Thus, the cosmological constant $\Lambda$ is identified with pressure.
This equation coincides with classical thermodynamics in which $H=E+PV$, and therefore $M$ is considered to be the enthalpy in this proposal. This model has been extended to include  extra variables and phase structures \cite{Dolan:2010ha,Kubiznak:2014zwa}, which further enrich the theory,
by assuming the fine tuning of extra variables so that a phase transition is induced.

However, virtually all of the previous extensions were based on the modifications of the first law.  Replacing the cosmological constant with pressure and discussing the evolution of a black hole alone (without taking environment into account) is incomplete, since the cosmological constant cannot be tuned freely like pressure. 
 For a self-consistent description, one has to describe how these variables actually change dynamically. Without the entropy of the environment,  a black hole entropy on its own is not a good indicator if a process is allowed or not.

 \begin{figure}
   \centering
\includegraphics[width=5cm]{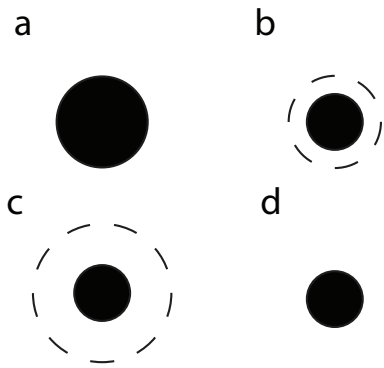}
\caption{A process in which  a black hole changes its state. Initially, the black hole is in the state a. Then it emits (for simplicity) a spherically symmetric wave, and the system changes its state to b. The dashed line represents emitted matter or field. The emitted matter keeps moving away from the black hole and the system changes its state to c. The matter completely escapes from a black hole and the state becomes  d. This represents an emission process ($a\rightarrow b\rightarrow c\rightarrow d$). An absorption process can be made by reversing the order of steps ($d\rightarrow c\rightarrow b\rightarrow a$).  
}
\label{wall}
\end{figure}

In general, an isolated black hole can change its own state by emitting (or absorbing) matter/energy or a field (e.g. in Fig.~\ref{wall}). Eventually, the emitted matter escapes to infinity and the black hole becomes an isolated object once again. Without taking into account the middle steps (b and c in Fig.~\ref{wall}), all the information about  the environment is neglected. A black hole entropy on its own can increase or decrease with no full thermodynamical  meaning. Here we want to study a complete system of the black hole plus environment, in the context of a realistic model where the cosmological constant changes its value. We will verify that the total entropy of the system always increases. 

\section{Black hole induced vacuum decay}

\begin{figure}
   \centering
\includegraphics[width=8cm]{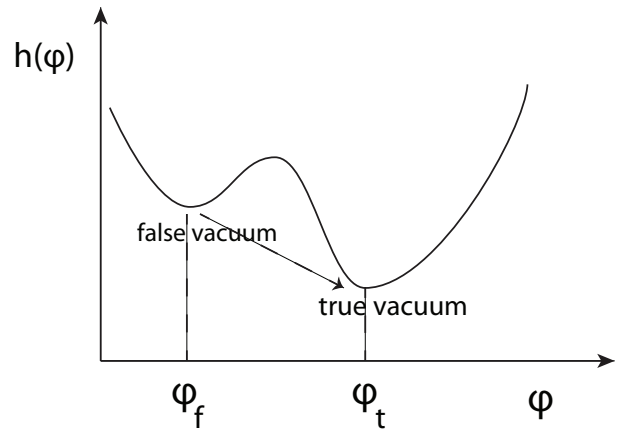}
\caption{The potential of the scalar field with two local minima. One is the false vacuum ($\varphi_f$) and the other one is the true vacuum ($\varphi_t$).
}
\label{potential-f}
\end{figure}

One of the ways to change the value of the cosmological constant is to tunnel from one de Sitter vacuum to another. [While we concentrate on a de Sitter space with positive cosmological constant, identical analysis can be done in anti-de-Sitter space with negative cosmological constant, what we will do at the end of the paper.]  This tunneling can be spontaneously triggered by the presence of the black hole, and has been well studied in the literature \cite{Gregory:2013hja,Burda:2015yfa}. For our purpose, we adopt a simple scalar field model where the non-zero vacuum energy of the scalar field, $\varphi$, plays the role of the cosmological constant.  The action of the model is 

\begin{equation}
S=\frac{1}{16\pi G} \int_M R \sqrt{-g} d^4x +\int_M \left[\partial_\mu \varphi\partial^\mu\varphi -h(\varphi) \right]\sqrt{-g} d^4 x ,
\end{equation}
where $R$ is the Ricci scalar, while $h(\varphi)$ is the scalar field potential. We set $\hbar=k=c=1$. To allow $\Lambda$ to change, we consider a potential like in Fig.~\ref{potential-f}.  The scalar field is initially stuck in the false vacuum, $\varphi_f$, and then decays to the true vacuum, $\varphi_t$. The cosmological constant goes from $\Lambda_+ =8\pi G h(\varphi_f) $ to  $\Lambda_- =8\pi G h(\varphi_t) $. Thus, the cosmological constant varies discretely in this model, and it can change only in one direction. A continuously varying $\Lambda$ model is possible by considering a quintessence-like potential, but for our purpose this discrete model will suffice.

To simplify the discussion, we focus on a spherically symmetric  metric as a background in which the scalar field $\varphi$ propagates
\begin{eqnarray}
ds^2 &=& -f(r) d t^2 +\frac{dr^2}{f(r)} +r^2 d\Omega \\
f(r) &=& 1-\frac{2GM_{\pm}}{r}-\frac{\Lambda_{\pm}r^2}{3} .
\end{eqnarray}
Here $M_+$ and $M_-$  are the values of the black hole mass before and after tunneling. In the process of the vacuum decay, a spherical bubble filled with new vacuum is formed. This bubble is enveloped with a spherical  domain wall separating two vacua.  Some of the original black hole mass/energy is invested into the phase transition, and its mass changes from $M_+$ to $M_-$.   Thus, some of the energy of the domain wall comes from the energy of the original black hole, so the situation is similar to the configuration b in Fig.~\ref{wall}.    

To study the dynamics of the bubble (domain wall), we use the thin wall approximation.  The energy density and wall tension depend on the potential  
\begin{equation}
\sigma=\left| \int_{\phi_f}^{\phi_t} (2h)^{1/2} d\phi \right|
\end{equation}
The absolute value mark is inserted to ensure the energy density is positive and physical. The cosmological constant takes values  $\Lambda_-$ and $\Lambda_+$ inside  and outside the bubble respectively,  and also  $f_-=1-\frac{2GM_{-}}{r}-\frac{\Lambda_{-}r^2}{3}$  and $f_+=1-\frac{2GM_{+}}{r}-\frac{\Lambda_{+}r^2}{3}$  inside  and outside the bubble respectively. The equation of motion of the wall can be obtained from the junction condition
\begin{equation}
\label{Lorentz-motion}
f_+ \dot{t}_+ - f_- \dot{t}_- =-4\pi G \sigma R ,
\end{equation}
with $\dot{x} =dx/d\lambda$ , where $\lambda$ is the proper time of an observer sitting on the wall enveloping the bubble. The wall is located at the radius $R$. {Combining this equation with the motion in the following coordinates}

\begin{equation}
f_{\pm} \dot{t}^2_{\pm} -\frac{\dot{R}_\pm^2}{f_\pm} =1,
\end{equation}
the equation of motion is simplified to 

\begin{equation}
\Big(\frac{\dot{R}}{R}\Big)^2=\bar{\sigma}^2-\frac{\bar{f}}{R^2}+\frac{(\Delta f)^2}{16R^4\bar{\sigma}^2} .
\end{equation}
Here, $\bar{\sigma}=2\pi G \sigma$, $\bar{f}=(f_-+f_+)/2$ and $\Delta f=f_+-f_-$. The exact solution represents a bounce, i.e. the bubble first contracts and then expands. To describe the expanding case, the solution is usually cut at the bounce.  Exactly at the bounce, the domain wall is generated with $\dot{R}_\pm=0$. Eq (\ref{Lorentz-motion}) implies 

\begin{equation}
\sqrt{f_-} -\sqrt{f_+} >0,
\end{equation}       
or 
\begin{equation}
M_+ - M_- > - \frac{R^{*3}}{6G} (\Lambda_+ -\Lambda_-).
\end{equation}
Here $R^*$ is the bubble radius of the bounce solution. $M_-$ must satisfy the following inequality
$ M_- < M_+ +\frac{R^{*3}}{6G} (\Lambda_+ -\Lambda_-)$.
This implies that the final black hole (after the tunneling) can be either heavier or lighter than the initial black hole. Since $\sigma$ cannot be $0$, the actual constraint is more stringent.

\begin{figure}
   \centering
\includegraphics[width=8cm]{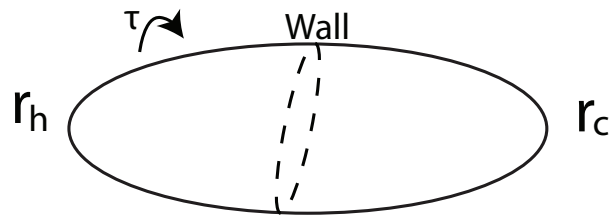}
\caption{A bubble (domain wall) with radius $R$ is generated in Euclidean space. The entire space is between the black hole horizon, $r_h$, and the cosmological horizon, $r_c$. The coordinate $\tau$ is periodic, and its period is the inverse temperature.    
}
\label{space}
\end{figure}

\section{The second law of thermodynamics for the whole system: black hole plus environment }

In order to study the thermodynamical properties of the wall, one must perform the Wick rotation, $t=-i\tau$. The action becomes the Euclidean action, and the metric becomes 

\begin{eqnarray}
ds^2 &=& f(r) d\tau^2 +\frac{dr^2}{f(r)} +r^2 d\Omega .
\end{eqnarray}
The domain wall equation of motion is 
\begin{equation}
-\Big(\frac{\dot{R}}{R}\Big)^2=\bar{\sigma}^2-\frac{\bar{f}}{R^2}+\frac{(\Delta f)^2}{16R^4\bar{\sigma}^2} .
\end{equation}
This equation can be written as 
\begin{equation}
\label{EOM}
\frac{\dot{R}^2}{2}+U=0 ,
\end{equation}
with 
\begin{eqnarray}
\label{potential}
2U&=& A R^2 -1+BR^{-1}+CR^{-4} \\
A&=&\bar{\sigma}^2 +\frac{1}{6}(\Lambda_+ +\Lambda_-)+\frac{(\Lambda_+-\Lambda_-)^2}{144\bar{\sigma}^2}\\
B&=&G(M_++M_-) +G\frac{(M_+-M_-)(\Lambda_+-\Lambda_-)}{12\bar{\sigma}^2}\\
C&=&G^2\frac{(M_+-M_-)^2}{4\bar{\sigma}^2} .
\end{eqnarray}

This equation is identical to that of a non-relativistic particle moving in a potential $U$ with total energy $0$. According to thermal quantum field theory, the period of $\tau$ is the inverse  temperature, which can be read out from the period of motion of the object. This period can be calculated from $\int dR/\dot{R}$. In the case of small oscillations around the bounce condition (i.e. $U=U'=0$), a simple harmonic oscillator approximation can be applied. Eq.~(\ref{EOM})  is then approximated as 
\begin{eqnarray}
&&\Delta E =\frac{\dot{R}^2}{2}+\frac{U''(R^*)}{2} \Delta R^2\\
&&U''(R)=A+BR^{-3}+10CR^{-6} ,
\end{eqnarray}
where $R^*$ is the radius of the bubble in the bounce solution and $\Delta R=R-R^*$. The period of $\lambda$ is
\begin{eqnarray}
\beta_\lambda=\frac{2\pi}{\sqrt{U''}} .
\end{eqnarray}

\begin{figure}
   \centering
\includegraphics[width=9cm]{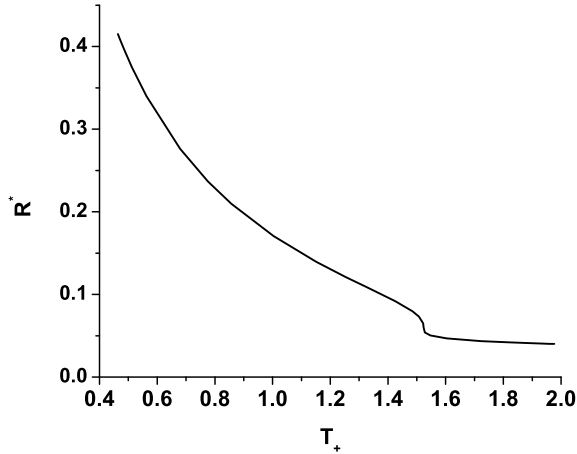}
\caption{ The relation between the bubble radius and its temperature. The bubble temperature gets lower as the radius is increasing. We set $M_+=0.02$, $\Lambda_+=3$, $\Lambda_-=0.03$ and $G=1$, and change the  black hole rest mass $M_-$.  
}
\label{radius}
\end{figure}

\begin{figure}
   \centering
\includegraphics[width=9cm]{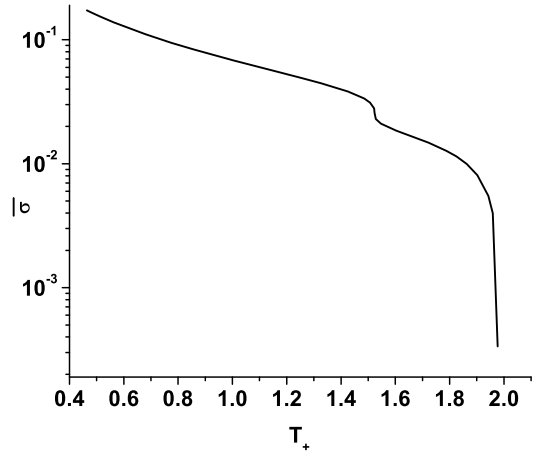}
\caption{The relation between the bubble wall's energy density ($\bar{\sigma}$) and its temperature ($T_+$). There is a sharp change at the point where $M_-=M_+$, which might indicate a phase transition of some sort. Physically, this represents the critical point where a black hole releases or absorbs energy from the vacuum decay.   We set $M_+=0.02$, $\Lambda_+=3$, $\Lambda_-=0.03$ and $G=1$, and change the black hole rest mass $M_-$.  
}
\label{sigma}
\end{figure}

The bubble temperature is  
\begin{eqnarray}
T_{\pm} =\frac{\sqrt{f_{\pm}(R)}}{\beta_\lambda} =\frac{\sqrt{U'' f_\pm(R)}}{2\pi} .
\end{eqnarray}
Here, $T_+$ and $T_-$ are the values of the temperature at the outer  and inner side of the wall respectively. In general $\bar{\sigma}$ is determined by the scalar field potential. Since we do not have a precise form of $f(\varphi)$ we keep it as a free parameter which will be determined by the initial and final states. We note from Eq.~(\ref{potential}) that $\bar{\sigma}$ cannot be $0$, which implies that there are no massless domain walls in this context. 
{Here $R^*$ and $\bar{\sigma}$ are obtained from $M_{\pm}$ and $\Lambda_\pm$.} The condition for $U=U'=0$ at $R^*$ is 
\begin{eqnarray}
&&2AR^{*6}-BR^{*3}-4C=0\\
&&6AR^{*3}-4R^*+3B=0 .
\end{eqnarray}
If $M_+ =M_-$, there is a special solution $R^*=3G M_+ $, where the value of $\bar{\sigma}$ determines whether this solution exists. The relation between $R^*$ and $T_+$ is shown in Fig.~\ref{radius}. Just like for black holes, larger radius implies lower temperature. There is a clear region of a sharp change at around $M_+=M_-$, which may be a sign of a phase transition.  The same transition can also be seen from the relation between the energy density and temperature, shown in Fig.~\ref{sigma}, where a very quick transition appears at around $M_-\approx M_+$. Physically, this transition may be explained as the critical point where a black hole releases or absorbs energy from the vacuum decay. 

\begin{figure}
   \centering
\includegraphics[width=9cm]{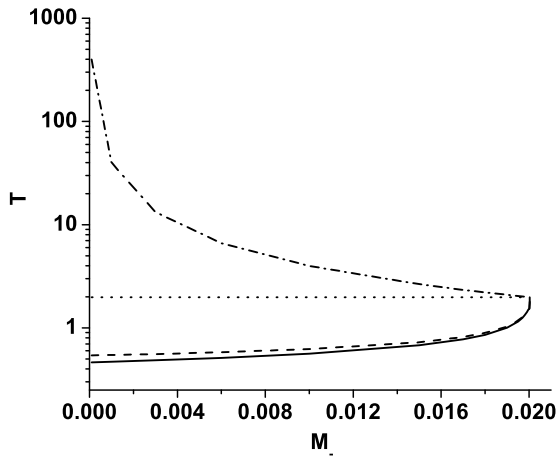}
\caption{The temperatures of the bubble and the black hole as functions of the black hole rest mass. The solid, dashed, doted and dot-dashed lines are the temperatures at the outer side of the bubble wall, inner side of the wall, initial black hole surface and final black hole surface respectively. We set $M_+=0.02$, $\Lambda_+=3$, $\Lambda_-=0.03$ and $G=1$, and change the black hole rest mass $M_-$.  
}
\label{temperature}
\end{figure}

\begin{figure}
   \centering
\includegraphics[width=9cm]{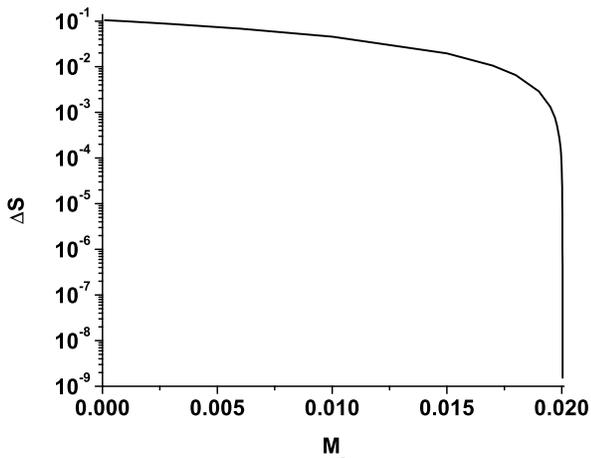}
\caption{The change in total  entropy of the system as a function of $M_-$ in de Sitter space. The entropy is always increasing. As long as  $M_-\lessapprox M_+$, $\Delta S$ reduces with $M_-$ slowly. The reduction becomes very quick when  $M_-\lessapprox M_+$. This implies that the black hole prefers to reduce its energy while triggering the vacuum decay.  We set $M_+=0.02$, $\Lambda_+=3$, $\Lambda_-=0.03$ and $G=1$, and change the black hole rest mass $M_-$.  
}
\label{entropy}
\end{figure}

\begin{figure}
   \centering
\includegraphics[width=9cm]{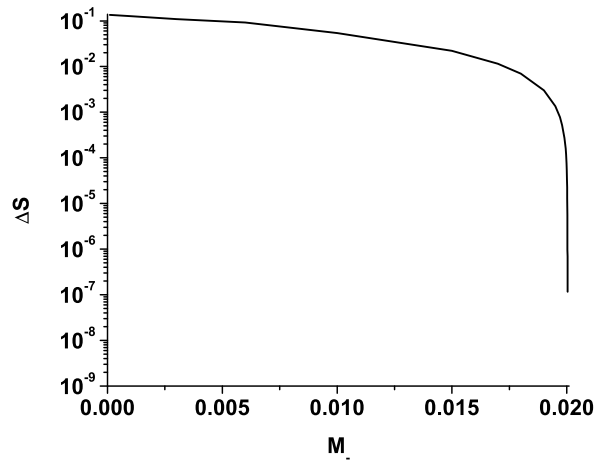}
\caption{The change in total  entropy of the system as a function of $M_-$ in anti de Sitter space. The situation is very similar to de Sitter space, i.e. the entropy is always increasing. As long as  $M_-\lessapprox M_+$, $\Delta S$ reduces with $M_-$ slowly. The reduction becomes very quick when  $M_-\lessapprox M_+$. This implies that the black hole prefers to reduce its energy while triggering the vacuum decay.  We set $M_+=0.02$, $\Lambda_+=-0.03$, $\Lambda_-=-3$ and $G=1$, and change the black hole rest mass $M_-$.  
}
\label{entropy-ads}
\end{figure}

Fig.~\ref{temperature} shows that the bubble has a lower temperature than the black hole horizon ($T_{Bh} = (1-\Lambda r_h^2)/(4\pi r_h)$) if the cosmological constant reduces its value in the process. The change in entropy can be calculated directly from the initial black hole, final black hole and the bubble's entropy as 
\begin{equation}\label{ds}
\Delta S=\pi r_{h-}^2+\frac{4\pi \sigma R^{*2}}{T_+}f_+(R^*)^{1/2}-\pi r_{h+}^2,
\end{equation}
where we set $G=1$. The outer (cosmological) horizon stays the same in this instantaneous process of tunneling, so its contribution cancels out. [Note that the situation is very similar in the anti-de-Sitter (AdS) case, though cosmological horizon is not involved.] We added the term $f_+(R^*)^{1/2}$ to compensate for the redshift. Fig.~\ref{entropy} and Fig. \ref{entropy-ads} show how entropy changes with $M_-$ in de-Sitter and anti-de-Sitter spaces respectively.  As expected, entropy always increases. As long as  $M_-\lessapprox M_+$, $\Delta S$ reduces with $M_-$ slowly. The reduction becomes very quick when  $M_-\gtrapprox M_+$.  We emphasize that the entropy of the bubble (second term in Eq.~(\ref{ds}) is crucial for providing that the total entropy of the system always increases. Without it, entropy could decrease, thus apparently violating the second law of thermodynamics. We note that this is the entropy created during the instanton process when wall has just been created, but has not been expanding yet. The expanding process also create entropy, which will be discussed in the next section.

\section{Other mechanisms which increase entropy}

Vacuum change, slow of abrupt, is always accompanied with particle production. A change in the value of the cosmological constant is no exception. In our case of the bubble production during tunneling, particles are generated during the nucleation and expanding phase. This particle production increases the entropy; however, it is not taken into account in the extended black hole first law of thermodynamics.     

First we consider the bubble nucleation phase, where particles are extracted from vacuum during the formation of  the domain wall (Fig.~\ref{wall} b). 
Though the vacuum decay involves gravitational effects, here we just want to clarify the essential physics of particle production and, therefore, 
in what follows we use the results from flat space.  In flat space, the spectrum of created particles is  obtained from the Bogoliubov transformation \cite{Yamamoto:1994te}
\begin{equation}
\label{creation}
n_p=\Big| \frac{B_p}{\pi A_p} \Big|^2 ,
\end{equation}
where $p$ is the particle  momentum and 
\begin{eqnarray}
A_p&=&\frac{\mu  I'_{ip}(x)K_{ip}(y)-\mu_0  I_{ip}(x)K'_{ip}(y)}{\mu  I'_{ip}(x)K_{ip}(x)-\mu  I_{ip}(x)K'_{ip}(x)}\\
B_p&=&\frac{-\mu  K'_{ip}(x)K_{ip}(y)+\mu_0  K_{ip}(x)K'_{ip}(y)}{\mu  I'_{ip}(x)K_{ip}(x)-\mu  I_{ip}(x)K'_{ip}(x)} ,
\end{eqnarray}
with, $x=\mu R$ and $y=\mu_0 R$. Here $\mu_0$ and $\mu$ are the masses of a particle in the false and true vacuum regions respectively. The functions  $I_\nu$ and $K_\nu$ are the modified Bessel functions of the first and second kind, while $i$ is the standard imaginary unit. Apparently, particles are created if $\mu\neq \mu_0$ and this instanton process is an irreversible process. The entropy can be calculated according to the Gibbons entropy formula  
\begin{equation}
S_G=-N k\sum_i p_i \log (p_i).
\end{equation}
where $N$ is the particle number and $p_i$ is the probability  for a particle to be in the  state $i$. These particles will also increase the entropy ensuring that the second law of thermodynamics is satisfied.  

Second, bubble expansion also creates particles  (Fig.~\ref{wall} c) \cite{Yamamoto:1994te, Maziashvili:2003sk, Maziashvili:2003kj}. The particle spectrum can be calculated again using Eq. \ref{creation} by replacing $A_p$ and $B_p$ with 
\begin{eqnarray}
A_p&=&\frac{\mu  H^{(1)}_{ip}{}'(x)H^{(2)}_{ip}(y)-\mu_0  H^{(1)}_{ip}(x)H^{(2)}_{ip}{}'(y)}{\mu_0  H^{(1)}_{ip}{}'(y)H^{(2)}_{ip}(y)-\mu_0  H^{(1)}_{ip}(y)H^{(2)}_{ip}{}'(y)}\\
B_p&=&e^{-p\pi}\frac{-\mu  H^{(1)}_{ip}{}'(x)H^{(1)}_{ip}(y)+\mu_0  H^{(1)}_{ip}(x)H^{(1)}_{ip}{}'(y)}{\mu_0  H^{(1)}_{ip}{}'(y)H^{(2)}_{ip}(y)-\mu_0  H^{(1)}_{ip}(y)H^{(2)}_{ip}{}'(y)} .
\end{eqnarray}
$H_\nu^{(1)}$ and $H_\nu^{(2)}$ are the first and second kind Hunkel functions.
Thus,  though the bubble expands like a classical object, the fields propagating in its background change their vacuum state which leads to particle production. The particle spectrum is not completely thermal (Eq.~(\ref{creation})), nevertheless, the entropy is increasing during this process.

If we repeat the same procedure  in which $\Lambda$ is increasing instead of decreasing, we will find that the bubble wall's energy density cannot be positive and is therefore unphysical.  

We can extend the evolution of the bubble to extreme cases. As the bubble wall approaches the cosmological horizon size, the whole horizon volume gets converted into a new vacuum with lower value of the cosmological constant, and the entropy will increase further. 
 The entropy increases by 
\begin{equation}
\pi r_{c-}^2 -\pi r_{c+}^2 ,
\end{equation}
where $r_{c\pm}$ is the radius of the cosmological horizon in the false and true vacuum.  
 
Since the entropy increases, this is also an irreversible process. However, once the wall crosses and leaves the cosmological horizon behind, it will take away entropy with it. The domain wall's entropy then disappears and does not leave (a classical) imprint on the stuff inside the cosmological horizon.

\begin{figure}
   \centering
\includegraphics[width=7cm]{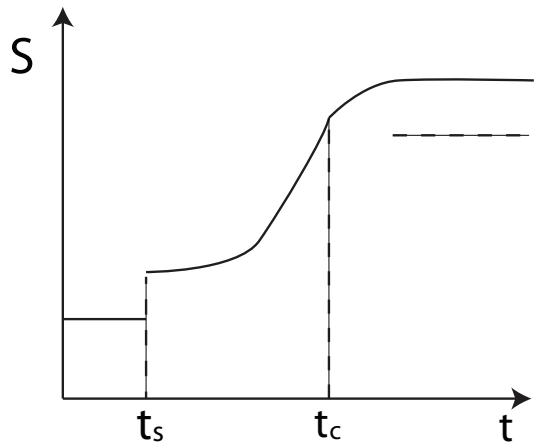}
\caption{ The schematic picture of the entropy evolution in time. The black hole triggers the vacuum decay through  instanton process at $t_s$. The entropy increases according to Eq.~(\ref{ds}), and in addition due to the entropy of particles created during the vacuum tunneling. The bubble wall produced during the tunneling keeps expanding and generating more entropy, since more particles are created during the expanding phase. At $t_c$ the wall grows to the size of the old cosmic horizon  and keeps pushing the horizon outward to its final size. Entropy is produced by both particle creation and the expansion of the cosmic horizon after $t_c$. Finally the wall leaves the new horizon and the entropy remains constant. The horizontal dashed line shows the entropy counted in the first law of the black hole thermodynamics, which is always lower than the actual entropy. 
}
\label{entropy-increasing}
\end{figure}

\section{Conclusions}

In this letter we have studied the extended black hole thermodynamics by discussing a realistic model in which the cosmological constant is variable. We have found that the vacuum decay triggered by an instanton has the temperature lower than the black hole horizon. During this process more entropy is produced than the change in the black hole entropy. 
The modified second law of thermodynamics for black holes only gives a complete picture provided the entropy of the environment is included in order to preserve the second law (see also \cite{Hu:2019lcy}).

To summarize, the entropy is increasing with time schematically as shown in Fig.~\ref{entropy-increasing}. If we neglect Hawking radiation, the total entropy is a constant at first. The black hole then triggers the vacuum decay through  instanton process at $t_s$. The entropy increases according to Eq.~(\ref{ds}), and in addition due to the entropy of particles created during vacuum tunneling. The bubble wall produced during the tunneling keeps expanding and generating more entropy, since more particles are created during the expanding phase. At $t_c$ the wall grows to the size of the old cosmic horizon (associated with the higher value of the cosmological constant)  and keeps pushing the horizon outward to its final size (associated with the lower value of the cosmological constant). Entropy is produced by both particle creation and the expansion of the cosmic horizon after $t_c$. Finally the wall leaves the new horizon and the entropy remains constant. The entropy created between $t_s$ and $t_c$ is not included in the first law of the extended black hole thermodynamics. The first law assumes the black hole entropy coming purely from the horizon size, which is always lower than the entropy shown here. The situation is very similar in the anti-de-Sitter case, though cosmological horizon is not involved.

Our analysis thus suggests that the environmental effects are crucial in the context of the extended black hole thermodynamics, and they call for
a more careful interpretation of the holographic principle \cite{holography}.
We note that such a sharpening of the idea of holography has been recently suggested in different contexts \cite{Freidel:2013jfa}, \cite{Almheiri:2020cfm}.
Finally, we note that our analysis is valid both for anti-de-Sitter (AdS) and de Sitter (dS) spaces, which we find significant, given their radically different
causal structures and holographic formulations \cite{ads}, \cite{ds}.

\begin{acknowledgments}
We thank D.~Kubiznak and R.~B.~Mann for comments.
D.C Dai is supported by the National Natural Science Foundation of China  (Grant No. 11775140). 
D. M. is supported by the Julian Schwinger Foundation and the Department of Energy ((under grant DE-SC0020262))).
D.S. is partially supported by the US National Science Foundation, under Grant No.  PHY-2014021.
\end{acknowledgments}


\begin{thebibliography}{99}




\bibitem{Bekenstein1972} J. D. Bekenstein, Lett. Nuovo Cim. 4 (1972) 737.
\bibitem{Bekenstein1973} J. D. Bekenstein, Phys. Rev. D 7 (1973) 2333.
\bibitem{Hawking1972} S. W. Hawking, Commun. Math. Phys. 25 (1972) 152.
\bibitem{Hawking1974}S. W. Hawking, Nature 248 (1974) 30.
\bibitem{Hawking1975} S. W. Hawking, Commun. Math. Phys. 43 (1975) 199.
\bibitem{Bardeen1973} J. M. Bardeen, B. Carter and S. W. Hawking, Commun. Math. Phys. 31 (1973) 161.



\bibitem{Henneaux1984} M. Henneaux and C. Teitelboim, Phys. Lett. B 143 (1984) 415.
\bibitem{Caldarelli:1999xj}
M.~M.~Caldarelli, G.~Cognola and D.~Klemm,
Class. Quant. Grav. \textbf{17}, 399-420 (2000)
[arXiv:hep-th/9908022 [hep-th]].



\bibitem{Sekiwa:2006qj}
Y.~Sekiwa,
Phys. Rev. D \textbf{73}, 084009 (2006)
[arXiv:hep-th/0602269 [hep-th]].


\bibitem{Kastor:2009wy}
D.~Kastor, S.~Ray and J.~Traschen,
Class. Quant. Grav. \textbf{26}, 195011 (2009)
[arXiv:0904.2765 [hep-th]].

\bibitem{Dolan:2010ha}
B.~P.~Dolan,
Class. Quant. Grav. \textbf{28}, 125020 (2011)
[arXiv:1008.5023 [gr-qc]].

\bibitem{Kubiznak:2014zwa}
D.~Kubiznak and R.~B.~Mann,
Can. J. Phys. \textbf{93}, no.9, 999-1002 (2015)
[arXiv:1404.2126 [gr-qc]];
D.~Kubiznak, R.~B.~Mann and M.~Teo,
Class. Quant. Grav. \textbf{34}, no.6, 063001 (2017)
[arXiv:1608.06147 [hep-th]].


 

\bibitem{Gregory:2013hja}
R.~Gregory, I.~G.~Moss and B.~Withers,
JHEP \textbf{03}, 081 (2014)
[arXiv:1401.0017 [hep-th]].

\bibitem{Burda:2015yfa}
P.~Burda, R.~Gregory and I.~Moss,
JHEP \textbf{08}, 114 (2015)
[arXiv:1503.07331 [hep-th]].


\bibitem{Yamamoto:1994te}
  K.~Yamamoto, T.~Tanaka and M.~Sasaki,
  Phys.\ Rev.\ D {\bf 51}, 2968 (1995)
  [gr-qc/9412011].
  
  

\bibitem{Maziashvili:2003sk}
  M.~Maziashvili,
  Mod.\ Phys.\ Lett.\ A {\bf 19}, 1391 (2004)
  [hep-th/0311263].
  
\bibitem{Maziashvili:2003kj}
  M.~Maziashvili,
  Mod.\ Phys.\ Lett.\ A {\bf 19}, 671 (2004)
  [hep-th/0311232].

\bibitem{Hu:2019lcy}
S.~Q.~Hu, Y.~C.~Ong and D.~N.~Page,
Phys. Rev. D \textbf{100}, no.10, 104022 (2019)
[arXiv:1906.05870 [gr-qc]].



\bibitem{holography}
G.~'t Hooft,  arXiv:gr-qc/9310026; 
L.~Susskind,
J. Math. Phys. \textbf{36}, 6377-6396 (1995)
[arXiv:hep-th/9409089 [hep-th]].

\bibitem{Freidel:2013jfa}
L.~Freidel,
Class. Quant. Grav. \textbf{32}, no.5, 055005 (2015)
[arXiv:1312.1538 [gr-qc]];
L.~Freidel and Y.~Yokokura,
Class. Quant. Grav. \textbf{32}, no.21, 215002 (2015)
[arXiv:1405.4881 [gr-qc]].


\bibitem{Almheiri:2020cfm}
A.~Almheiri, T.~Hartman, J.~Maldacena, E.~Shaghoulian and A.~Tajdini,
Rev. Mod. Phys. \textbf{93}, no.3, 035002 (2021)
[arXiv:2006.06872 [hep-th]].

\bibitem{ads}
J.~M.~Maldacena,
Adv. Theor. Math. Phys. \textbf{2}, 231-252 (1998)
[arXiv:hep-th/9711200 [hep-th]];
S.~S.~Gubser, I.~R.~Klebanov and A.~M.~Polyakov,
Phys. Lett. B \textbf{428}, 105-114 (1998)
[arXiv:hep-th/9802109 [hep-th]].
E.~Witten,
Adv. Theor. Math. Phys. \textbf{2}, 253-291 (1998)
[arXiv:hep-th/9802150 [hep-th]].


\bibitem{ds}
A.~Strominger,
JHEP \textbf{10}, 034 (2001)
[arXiv:hep-th/0106113 [hep-th]].
E.~Witten,
[arXiv:hep-th/0106109 [hep-th]];
V.~Balasubramanian, P.~Horava and D.~Minic,
JHEP \textbf{05}, 043 (2001)
[arXiv:hep-th/0103171 [hep-th]];
V.~Balasubramanian, J.~de Boer and D.~Minic,
Phys. Rev. D \textbf{65}, 123508 (2002)
[arXiv:hep-th/0110108 [hep-th]];
V.~Balasubramanian, J.~de Boer and D.~Minic,
Class. Quant. Grav. \textbf{19}, 5655-5700 (2002)
[arXiv:hep-th/0207245 [hep-th]].


\end{thebibliography}
\end{document}